\documentclass[3p]{elsarticle}

\usepackage{epsfig}

\usepackage{amssymb}
\usepackage{amsmath}
\usepackage{amsthm}
\usepackage{subfigure}
\usepackage{xcolor}
\usepackage{float}
\usepackage{todonotes}

\newcommand{\bfB}{{\bf B}}%
\newcommand{\bfF}{{\bf F}}%
\newcommand{\bfn}{{\bf n}}%
\newcommand{\bfu}{{\bf u}}%
\newcommand{\bfx}{{\bf x}}\newcommand{\bfX}{{\bf X}}%
\newcommand{\bfsigma}{\boldsymbol{\sigma}}%
\newfont{\tenbfit}{cmmib10}%
\newfont{\svnbfit}{cmmib8}%
\newfont{\tenbfsl}{cmbxti10}
\newfont{\mmit}{cmmi10}
\newfont{\smit}{cmmi9}
\newfont{\bfMit}{cmmi5}
\newfont{\tenbbb}{msbm10}%
\newfont{\svnbbb}{msbm8}%
\newfont{\tenssit}{cmssqi8 at 10pt}%
\newfont{\svnssit}{cmssqi8 at 7pt}%
\newfont{\gothic}{eufm10}%
\newfont{\sgothic}{eufm7}%

\newcommand{\pards}[2]{\mbox{$\dfrac{\partial #1}{\partial {#2 }}$}}

\newcommand{\Blj}{\mbox{$\Big[\kern-0.275em\Big[$}}
\newcommand{\Brj}{\mbox{$\Big]\kern-0.275em\Big]$}}

\newcommand{\X}{\bfX}
\newcommand{\x}{\bfx}

\begin{document}

\begin{frontmatter}

\title{\textit{Push} and \textit{Pull}: Elastic Interaction Between Pressurized Spherical Cavities in Nonlinear Elastic Media}
\author[inst1]{Ali Saeedi}
\author[inst1]{Mrityunjay Kothari\corref{cor1}}
\address[inst1]{Department of Mechanical Engineering, College of Engineering and Physical Sciences, University of New Hampshire, Durham, NH 03824, USA}

\cortext[cor1]{Corresponding Author. Email: Mrityunjay.Kothari@unh.edu }

\begin{abstract}
Elastic interaction of pressurized spherical cavities embedded in a three-dimensional hyperelastic medium is computationally analyzed. 
Using finite element analysis across several positive and negative pressure scenarios, we calculate the system's potential energy and configurational driving force for neo-Hookean, Mooney-Rivlin, and Arruda-Boyce material models. 
Our results show that while the interaction is always attractive for negative pressures, a non-monotonic energy landscape emerges for positive pressures above a critical value. 
In this regime, cavities attract at close range and repel when further apart. 
The critical separation distance for this transition is shown to be dependent on the material's strain-stiffening parameters. 
These findings are consolidated into phase diagrams, providing a clear map of interaction behaviors.

\endkeyword
\end{abstract}

\end{frontmatter}

\makeatother


\section{Introduction}
\label{sec:intro}

\noindent This work investigates the elastic interaction between two pressurized cavities embedded in an infinite elastic medium.
This is the simplest conceivable problem to study inter-cavity elastic interaction effects when the cavities are \textit{not} sufficiently far enough from each other to ignore their interaction.
Extensive prior work has over the last century contributed to establishing that when the medium is linear elastic, the interaction between pressurized cavities is always attractive.
In other words, the total potential energy of the system, when comparing two equi-sized, equi-pressurized cavities, decreases as the distance between the cavities decreases \cite{willis1969, lidiard1968gas, Eshelby1955, eshelby1957elastic, chalon2003, Willis1975, johnson1982integral,Choi1985, jeffery1921ix}.

However, there is currently a lack of understanding of such interactions when the medium is non-linear elastic.
To address this, in this work, we computationally study the elastic interaction of two equi-sized pressurized spherical cavities embedded within an infinite hyperelastic medium. 
The analysis is founded on the calculation of the system's total potential energy and the corresponding \textit{configurational} driving force, which quantifies the tendency for the cavities to \textit{attract} or \textit{repel}. 
We explore this interaction across a range of inter-cavity distances and internal pressures, considering both positive (expansive) and negative (contractile) pressure scenarios, 
which is critical for understanding diverse phenomena ranging from liquid-liquid phase separation in elastic networks \citep{Kothari2020, kothari2023crucial, style2018liquid} to the contractile forces exerted by migrating cells \citep{legant2010measurement} and the morphogenesis of bacterial biofilms \citep{zhang2021morphogenesis,nijjer2023biofilms}.
To capture a comprehensive picture of non-linear material effects, we employ three widely used isotropic, incompressible hyperelastic models: neo-Hookean (NH), Mooney-Rivlin (MR), and Arruda-Boyce (AB) \citep{treloar1943elasticity, rivlin1948large, mooney1940theory, Arruda1993}, and we specifically examine the role of strain-stiffening.

 Our results reveal notable departures from linear elastic predictions.
 While the interaction remains purely attractive for all cases of negative pressure, we identify a critical pressure threshold for positive pressures above which the interaction can show both attraction and repulsion depending on the inter-cavity distance.
 Below a critical separation distance, the cavities attract, and beyond this distance, they repel each other. 
 This transition from attraction to repulsion is shown to be dependent on the specific hyperelastic model and can be tuned by the material's strain-stiffening characteristics. These findings are presented through analyses of the potential energy landscapes and phase diagrams illustrating the attractive and repulsive regimes, offering new insights into the complex mechanics of defect interactions in soft solids.

The problem setup and analysis approach is described in section 2, followed by results and discussion in section 3, and conclusion in section 4.


\section{Problem Setup and Analysis}
\subsection{Governing  Equations}
\begin{figure}[h]
\centering

  \includegraphics[scale = 0.7]{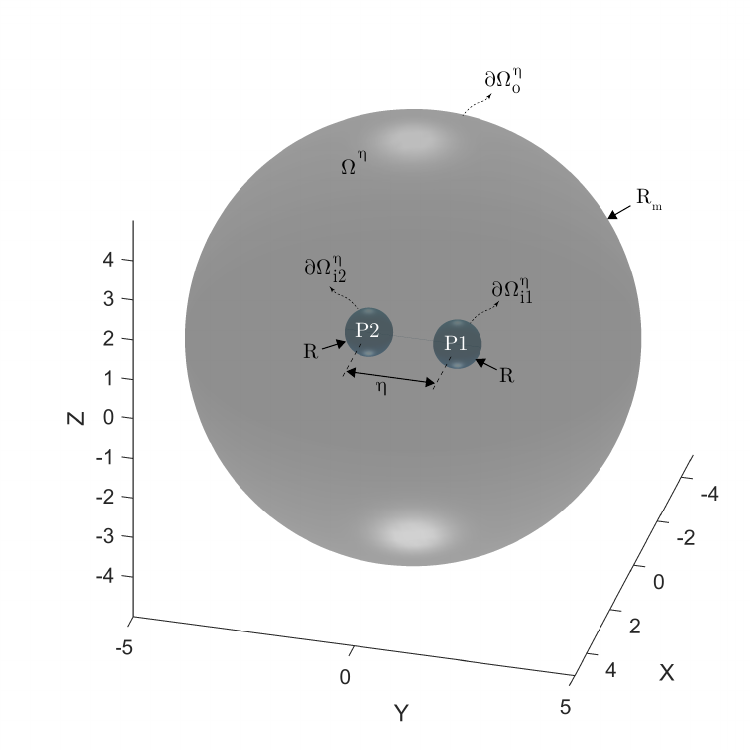}\\
  \caption{Schematic of the two-cavity system in a hyperelastic medium under constant internal pressures  $P_1$ and $P_2$.}
  \label{fig:Geometry}
\end{figure}

\noindent We consider two equi-sized spherical cavities of undeformed radius $R$ in a spherical domain that is sufficiently large to mimic an infinite medium.
The outer boundaries of the domain are traction-free.
The center-to-center distance between the cavities in their undeformed (reference) configuration is denoted by $\eta$.
The cavities themselves are internally pressurized with pressures $P_1$ and $P_2$ respectively, such that the traction boundary conditions are
\begin{equation}
\begin{cases} 
\displaystyle \bfsigma \bfn = -P_1\bfn \text{ for } \bfx \in \partial \Omega^{\eta}_{i1},\\
\displaystyle \bfsigma \bfn = -P_2\bfn \text{ for } \bfx \in \partial \Omega^{\eta}_{i2},\\
\displaystyle \bfsigma \bfn = \mathbf{0} \text{ for } \bfx \in \partial \Omega^{\eta}_{o}
\end{cases}
\end{equation}
where $\bfsigma$ is the Cauchy Stress tensor, $\bfn$ is the outward normal vector to the surface (points inside the cavities), $\partial\Omega^{\eta}_{i1}$ and $\partial\Omega^{\eta}_{i2}$ are the boundaries of the two cavities, and $\partial \Omega^{\eta}_{o}$ is the outer boundary of the domain, all in deformed (current) configuration.
Note that in this setup, the pressures are prescribed in the current configuration.
Material points in the body in the reference configuration are denoted by $\X$, and the corresponding points in the deformed configuration are denoted by $\bfx$, which gives the displacement field as $\bfu =\bfx -\bfX$.
The schematic of the problem setup is depicted in Fig. \ref{fig:Geometry}.

The geometry of the problem is indexed by $\eta$ to denote a sequence of configurations where the cavities are at varying distances.
The interaction between the cavities is expected to be a function of $\eta$; whether the interaction is attractive (ie. the equilibrium potential energy of the system decreases with decreasing $\eta$) or repulsive (ie. the equilibrium potential energy of the system increases with decreasing $\eta$) can be conveniently determined by the analyzing the \textit{driving force} for the problem.

To make concrete the notion of driving force, we first construct the total potential energy of the system in a three-dimensional setting as,

\begin{equation}
	\Pi^{\eta}[\bfu] = \int_{\Omega^{\eta}}\psi \mathrm{\ d\Omega} - P_1 \Delta V_1 - P_2 \Delta V_2, \label{PE}
\end{equation} 
where $\bfu$ is the displacement field, $\psi \equiv \psi(\bfF)$ is the strain energy density in the deformed configuration, $\bfF = \pards{\bfx}{\bfX}$ is the deformation gradient, $P_1$ and $P_2$ are the pressures on the inner boundaries $\partial \Omega^{\eta}_{i1}$ and $\partial \Omega^{\eta}_{i2}$ in the deformed configuration of the domain, and $\Delta V_1$ and $\Delta V_2$ are the changes in volumes of the cavities due to deformation.
For every admissible $\eta$, the equilibrium displacement field and the associated potential energy are denoted by $\bfu^{\eta}_{\rm{eqm}}$ and $\Pi^{\eta}_{\rm{eqm}} := \Pi^{\eta}[\bfu^{\eta}_{\rm{eqm}}]$, respectively, with both $\bfu^{\eta}_{\rm{eqm}}$ and $\Pi^{\eta}_{\rm{eqm}}$ being implicitly dependent on center-to-center distance of cavities, $\eta$.
The driving force $\mathcal{F}$ is then defined as \cite{lidiard1968gas, saeedi2025elastic},
\begin{equation}
	\mathcal{F} = -\frac{d\Pi^{\eta}_{\rm{eqm}}}{d\eta}. \label{driving_force}
\end{equation}
A positive driving force implies that the interaction is repulsive and vice versa.
Furthermore, we denote non-dimensionalized potential energy by
\begin{equation}
	\tilde{\Pi}^{\eta}_{\rm{eqm}} = \Pi^{\eta}_{\rm{eqm}}/|\Pi^{\infty}_{\rm{eqm}}|,
\end{equation}
where $\Pi^{\infty}_{\rm{eqm}}$ is the potential energy in the limiting case, when the cavities are infinitely far away and, therefore, there is no interaction between them.
In the same vein, the non-dimensionalized driving force is given as
\begin{equation}
	\tilde{\mathcal{F}} = -\frac{d\tilde{\Pi}^{\eta}_{\rm{eqm}}}{d(\eta/R)} = \frac{\mathcal{F}R}{|\Pi^{\infty}_{\rm{eqm}}|}.
\end{equation}
The definition of configurational force depends on the sequence of problems (which are indexed by $\eta$) that we choose to compare, which in turn is usually motivated from the physical problem being modeled.
In this work, we have constructed the sequence such that all the problems in the sequence have the same deformed pressures, $P_1$ and $P_2$.
This setup can represent problems where pressure is actively maintained, such as in biological systems, phase separation, or externally maintained.
Such a setting would not, however, be applicable to processes where the pressure and volume are coupled together, such as in expansion of gas bubbles.
\subsection{Hyperelastic Material Models in This Study}\label{sec:2.2}
\noindent The study focuses on three hyperelastic material models  that are commonly used to model a wide range of isotropic, incompressible, nonlinear elastic materials: neo-Hookean (NH) , Mooney-Rivlin (MR), and Arruda-Boyce (AB) models. 
These materials models are characterized by their strain energy functions, which are defined in terms of the left Cauchy-Green deformation tensor, $\bfB=\bfF\bfF^\top$, and its two invariants $\mathbb{I}_1 \text{ and } \mathbb{I}_2$,
\begin{align}
	\mathbb{I}_1 =& \operatorname{tr}(\bfB)\\
	\mathbb{I}_2 =&\frac{1}{2}\left((\operatorname{tr}(\mathbf{B}))^2 - \operatorname{tr}(\mathbf{B}^2)\right)
\end{align}
The respective strain energy densities for the chosen material models are
\begin{gather}
    \psi_{\text{NH}} = \frac{\mu}{2}(\mathbb{I}_1 - 3) \quad\label{eq:W_NH}\\
\psi_{\text{MR}} = \alpha\frac{\mu}{2} (\mathbb{I}_1 - 3) + \left(1-\alpha\right)\frac{\mu}{2} (	\mathbb{I}_2 - 3)\quad  \label{eq:W_MR}\\
\psi_{\text{AB}} = C_{2} \Big[\frac{1}{2}(	\mathbb{I}_1 - 3) + \frac{1}{20 \lambda_m^2} (\mathbb{I}_1^2 - 9) + \frac{11}{1050 \lambda_m^4} (\mathbb{I}_1^3 - 27) + \frac{19}{7000 \lambda_m^6} (\mathbb{I}_1^4 - 81) + \frac{519}{673750 \lambda_m^8} (	\mathbb{I}_1^5 - 243)\Big] \quad 
    \label{eq:W_AB}
\end{gather}
where $\mu$ is the shear modulus, and the parameters $0 < \alpha < 1$ and $\lambda_m$ control the degree of strain stiffening in MR and AB models \cite{Arruda1993}, respectively. 
The consistency condition for the incompressible AB model is \cite{Arruda1993} 
\begin{gather}
    \mu = C_2 \left( 1 + \frac{3}{5 \lambda_m^2} + \frac{99}{175 \lambda_m^4} + \frac{513}{875 \lambda_m^6} + \frac{42039}{67375 \lambda_m^8} \right).
\end{gather}


\subsection{Finite Element Modeling}
\begin{figure}[ht]
\centering
    \includegraphics[scale = 0.25] {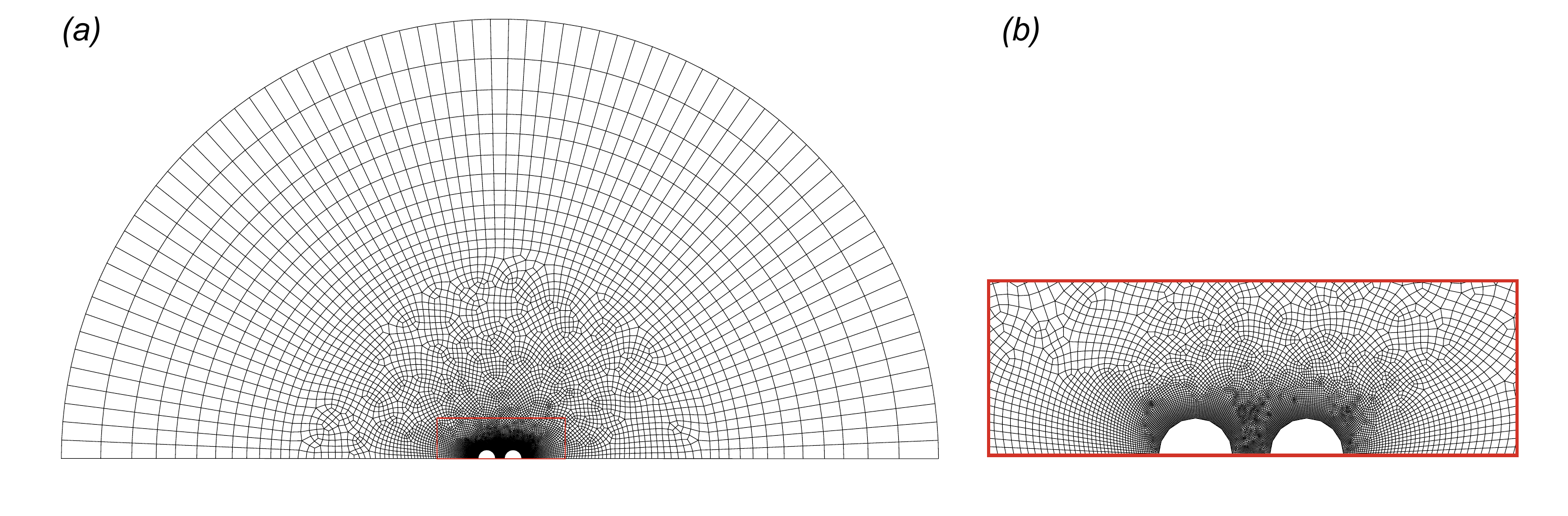}
\caption{A representative mesh employed in ABAQUS. (a) Entire domain and (b) locally refined region for $\eta/R=3$ using 11398 CAX4RH elements.}
\label{dicretized_geometry}
\end{figure}

\noindent We carry out the computational simulations using a commercial finite element software ABAQUS \cite{Abaqus}.
Since the problem is axisymmetric, we reduce the three-dimensional geometry, with domain radius 5 units and cavity radii 0.1 units, to an axisymmetric two-dimensional geometry, and discretize it using axisymmetric, bilinear quadrilateral hybrid elements with reduced integration (CAX4RH; a representative mesh is shown in Fig. \ref{dicretized_geometry}).

For each material model, a sequence of meshes are constructed for $\eta/R \in [2.2, 20]$ with interval of $\eta/R = 0.2$.
Each of these models is then simulated for a range of pressures, which is described in detail in the Results and Discussion section.
From each simulation, we obtain all the displacements and the total strain energy over the entire domain.
These data are post-processed to calculate the deformed volume of the cavities, and subsequently the potential energy for each run.

The potential energy of a neo-Hookean medium containing a single cavity is calculated and verified with its analytical solution in Appendix \ref{appendix:single_cavity}.
The actual simulation data, prior to fitting to compute derivatives, is reported in Appendix B.
Mesh-independence study is carried out and reported in Appendix C.

\section{Results and Discussion}
\subsection{Equal Pressures}
\noindent The primary focus of the analysis is on equi-pressurized cavities.
In this subsection, we consider cases where cavities are subjected to equal positive and equal negative pressures.
\subsubsection{Linear Elastic Medium}
\noindent We start by presenting the results for a linear elastic case to set the baseline for comparison with the nonlinear elastic materials.
\begin{figure}[H]
\centering
	 \includegraphics[scale = 0.25]{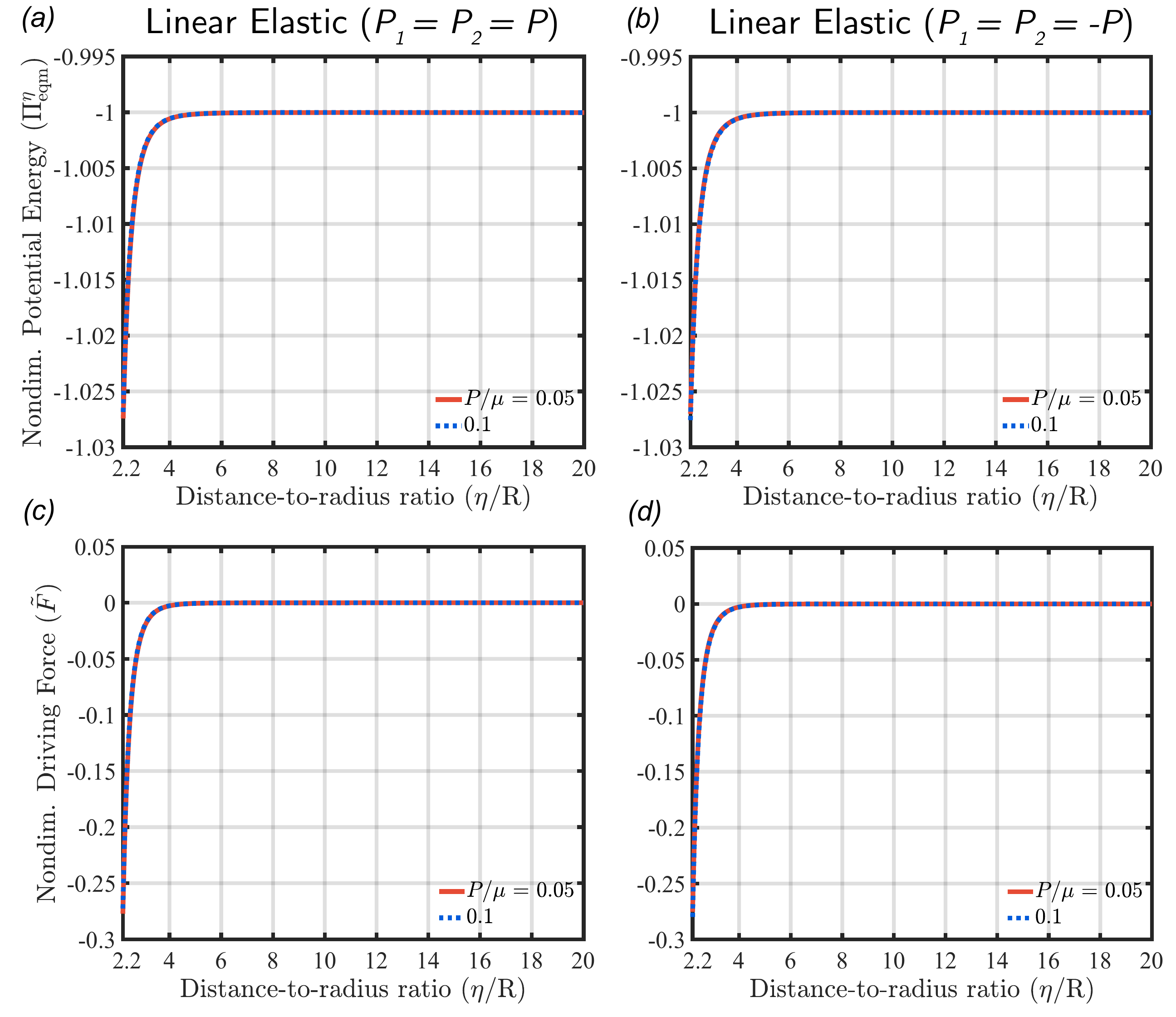}
\caption{\small Potential energy (a, b) and driving force (c, d) for equal positive and equal negative pressures, respectively, in linear elastic medium. $\mu=20$ was chosen for ABAQUS simulations.}
    \label{fig:PE_DF_LE}
\end{figure}
\noindent Figure \ref{fig:PE_DF_LE} shows the non-dimensionalized potential energy and driving force; here the non-dimensionalization is by $|\Pi^{\infty}_{\rm{eqm}}|$, which is the potential energy of the two-cavity system when the cavities are infinitely far apart. 
In other words, $\Pi^{\infty}_{\rm{eqm}}$ is twice the potential energy of a single pressurized cavity case.
Results show that the potential energy increases monotonically with $\eta$.
This shows that the interaction between cavities is \textit{always} attractive for equal pressures.
It follows that the driving force, which is the negative derivative of potential energy with $\eta$, is always negative, implying attraction between the cavities.
Notably, this trend is the same as observed in two-dimensional plane-strain results for cylindrical cavities \cite{saeedi2025elastic}.
Since in linear elasticity, strain energy and, consequently, potential energy are proportional to $P^2$ they show the same attractive trend for both equal negative and equal positive pressures. 
To ensure consistency with small deformation assumption in linear elasticity, we limit the range of pressure to $P/\mu\ll 1$. 

\subsubsection{Non-linear Elastic Medium}
\begin{figure}[H]
\centering
      \includegraphics[scale = 0.24]{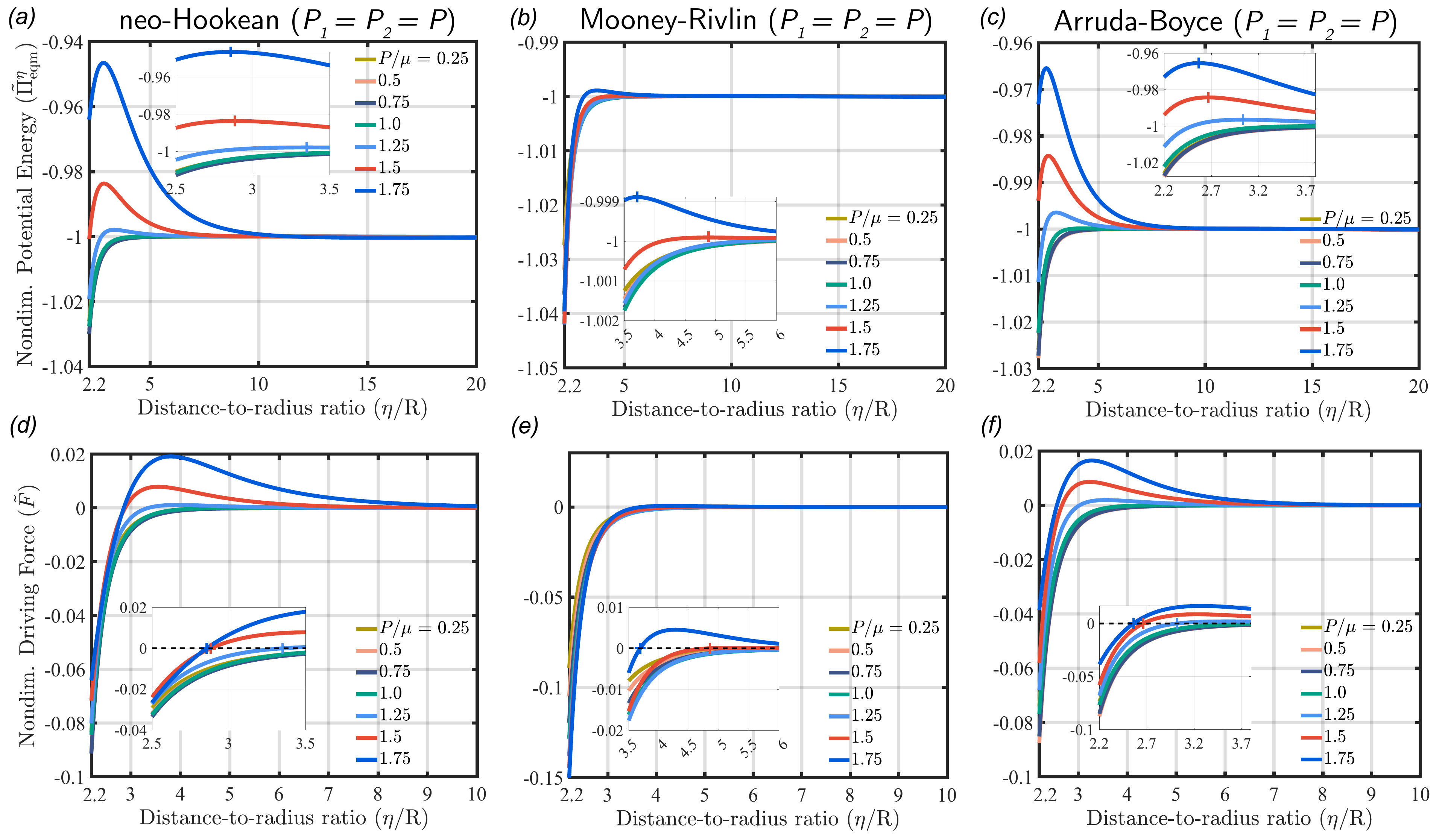}
\caption{\textbf{Positive pressure case.} Potential energy (a-c) and driving force (d-f) for neo-Hookean, Mooney-Rivlin, and Arruda-Boyce models. For all models, a value of $\mu=20$ was chosen for ABAQUS simulations. For Mooney-Rivlin, $\alpha = 0.5$, and for Arruda-Boyce, $\lambda_m=2$, were chosen as the constitutive model parameters.}    \label{fig:pos_pressure}
\end{figure}

\begin{figure}
\centering
       \includegraphics[scale = 0.25] {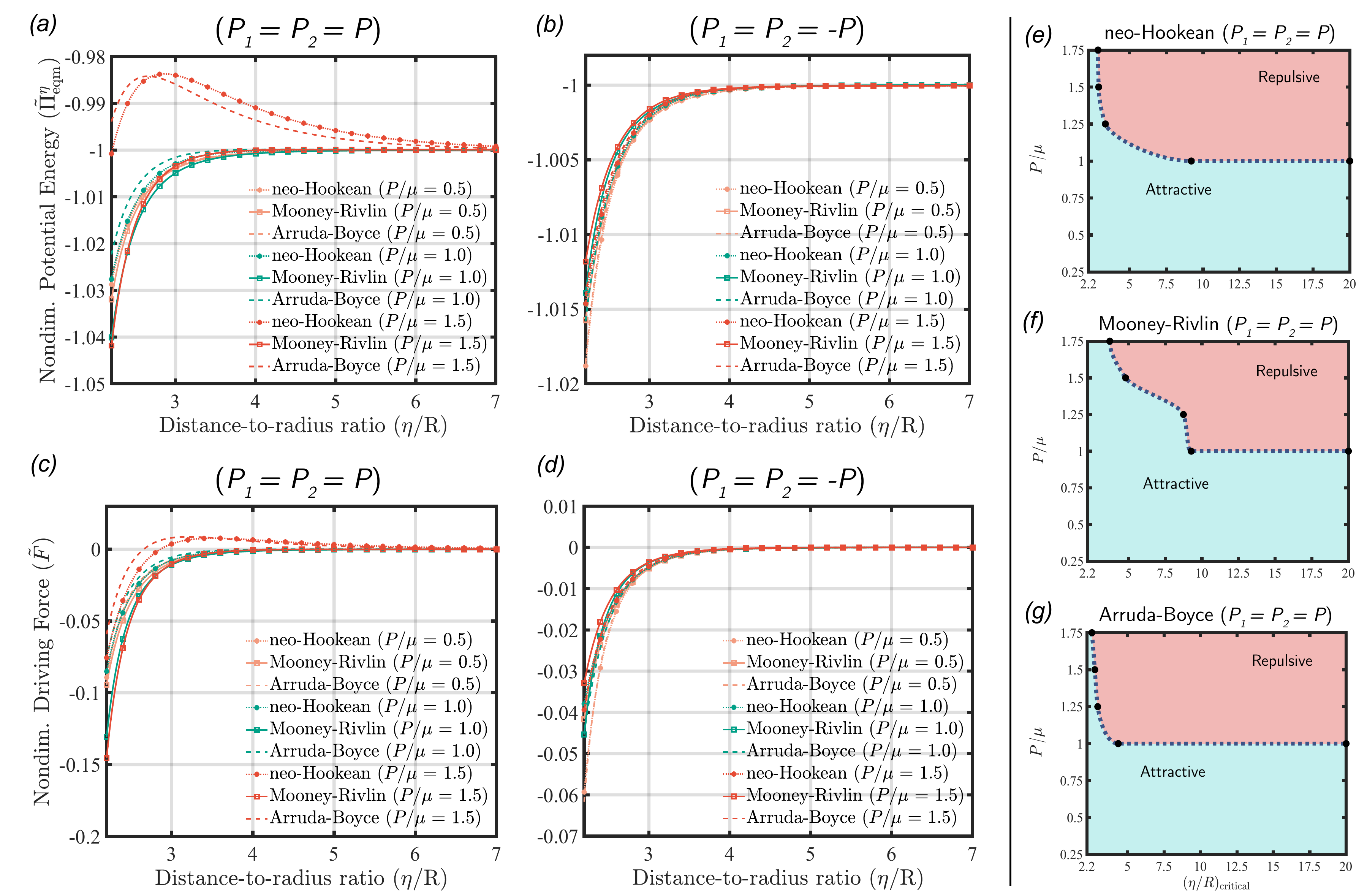}
\caption{Comparison of potential energy (a, b) and driving force (c, d) for nonlinear elastic constitutive models. For all models, a value of $\mu=20$ was chosen for ABAQUS simulations. For Mooney-Rivlin, $\alpha = 0.5$, and for Arruda-Boyce, $\lambda_m=2$, were chosen as the constitutive model parameters. Phase diagrams for neo-Hookean (e), Mooney-Rivlin (f), and Arruda-Boyce (g) models are shown for positive pressure case.}
\label{fig:PE_DF_comp_PD} 
\end{figure}

\noindent Figure \ref{fig:pos_pressure} illustrates the potential energy and driving force for equal positive pressures as a function of non-dimensionalized center-to-center distance between cavities ($\eta/$R) and for a range of non-dimensionalized pressures ($P/\mu$). 
At lower pressures, $P/\mu\lesssim1$, all models qualitatively follow the same trend as linear elasticity ie. the potential energy increases with increasing separation between cavities, and the interaction is attractive, as can be conveniently interpreted from  the negative driving forces.
However, at higher pressures, $P/\mu\gtrsim1$, a notable non-monotonicity emerges in the potential energy.
Below a critical value of separation, $(\eta/R)_{\rm{critical}}$, the cavities are attracted to each other, and for separation greater than this critical value, cavities repel each other.
This non-monotonicity is more pronounced in neo-Hookean and Arruda-Boyce models than in Mooney-Rivlin.
$(\eta/R)_{\rm{critical}}$ varies with pressure and material models as shown in a phase diagram in Fig. \ref{fig:PE_DF_comp_PD}.

\begin{figure}[H]
\centering
      \includegraphics[scale = 0.24]{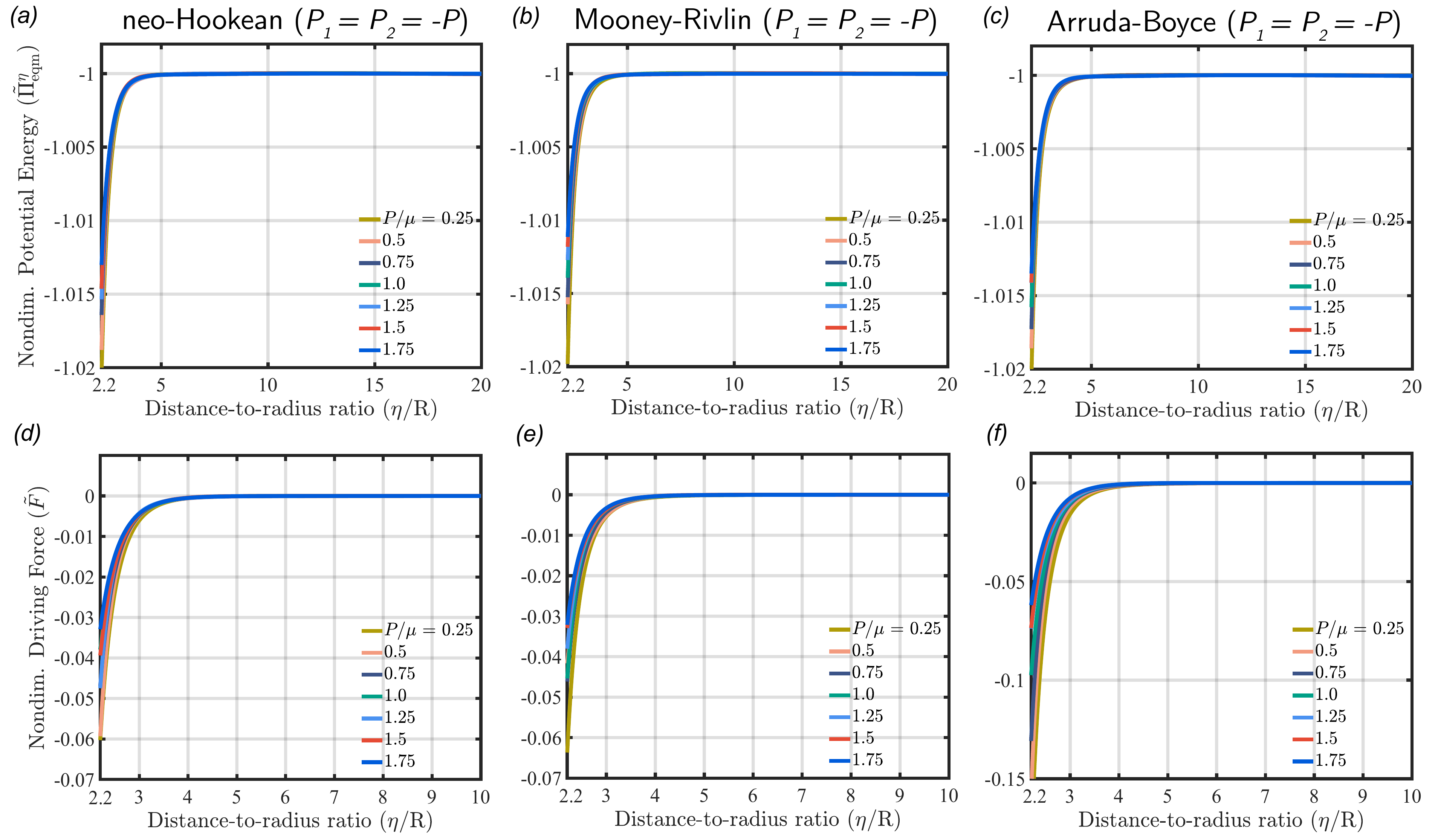}
\caption{\textbf{Negative pressure case.} Potential energy (a-c) and driving force (d-f) for neo-Hookean, Mooney-Rivlin, and Arruda-Boyce models. For all models, a value of $\mu=20$ was chosen for ABAQUS simulations. For Mooney-Rivlin, $\alpha = 0.5$, and for Arruda-Boyce, $\lambda_m=2$, were chosen as the constitutive model parameters.}    \label{fig:neg_pressure}
\end{figure}

In contrast to equal positive pressure, for equal negative pressure, the potential energy is monotonically increasing for the entire range of pressures tested, and for all the three material models.
The driving force is always negative, implying attractive interactions between the cavities (Fig. \ref{fig:neg_pressure}).


Figure \ref{fig:PE_DF_comp_PD} shows the comparison between the different constitutive models for representative values of $\alpha =0.5$ and $\lambda_m = 2$. 
There is a notable difference in the behavior of Mooney-Rivlin model when compared to neo-Hookean and Arruda-Boyce models.
A potential source of this difference is that while Mooney-Rivlin model depends on both $\mathbb{I}_1$ and ${\mathbb{I}_2}$, neo-Hookean and Arruda-Boyce models depend only on  $\mathbb{I}_1$.

\begin{figure}
\centering
       \includegraphics[scale = 0.25] {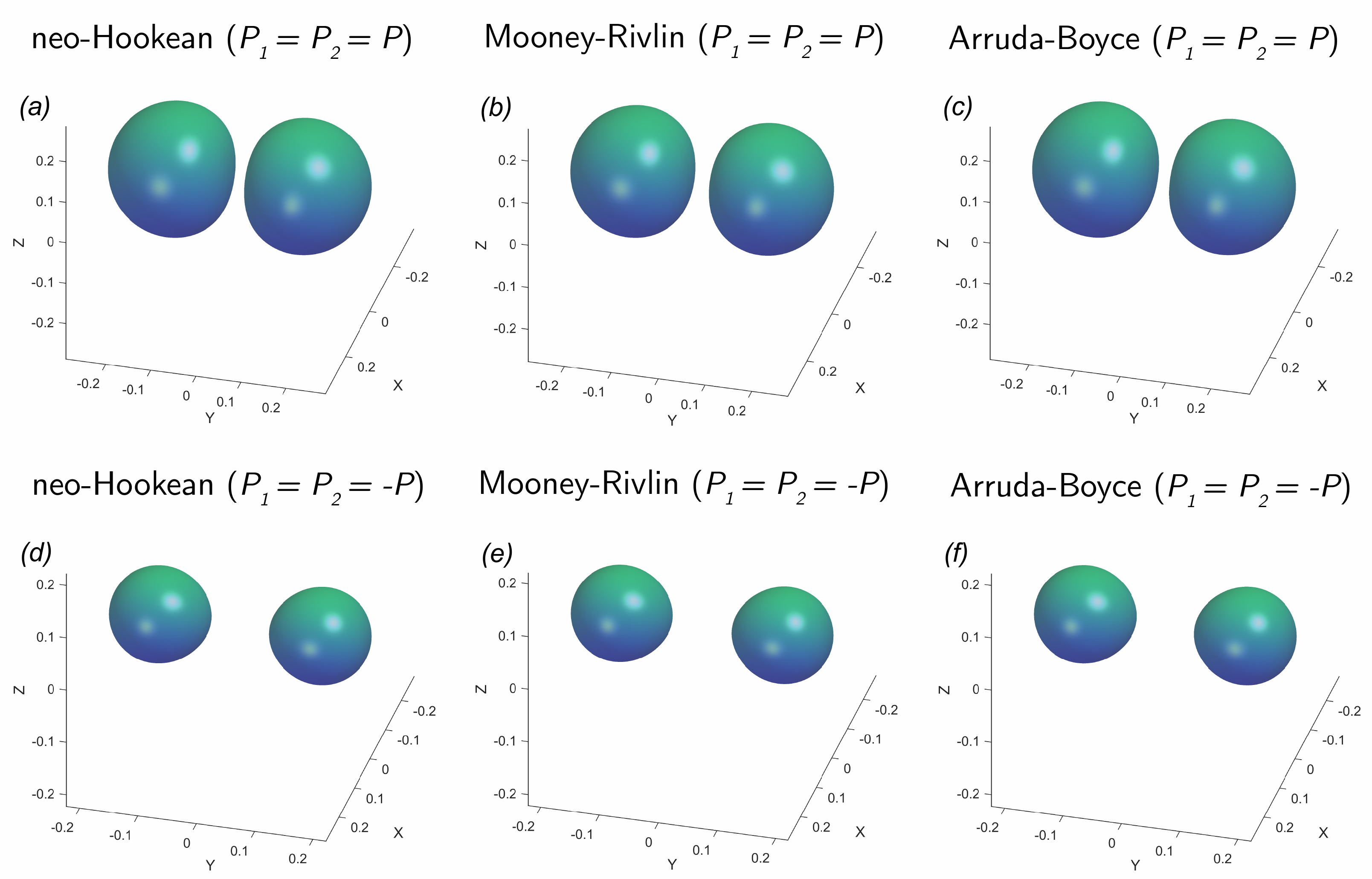}
\caption{Deformed shape of cavities for $\eta/R$=2.8 and $P/\mu$=1.0 with equal positive (a-c) and equal negative (d-f) pressures for neo-Hookean (a, d), Mooney-Rivlin (b, e), and Arruda-Boyce (c, f) models. For all models, a value of $\mu=20$ was chosen for ABAQUS simulations. For Mooney-Rivlin, $\alpha = 0.5$, and for Arruda-Boyce, $\lambda_m=2$, were chosen as the constitutive model parameters.}
\label{fig:def_crit_I_II}
\end{figure}

Representative deformed shapes of cavities for both equal positive and equal negative pressures are shown in Figure \ref{fig:def_crit_I_II} for $\eta/R$=2.8 and pressure of $P/\mu$=1.0.

\subsubsection{Effect of Strain Stiffening}
\begin{figure}
\centering
       \includegraphics[scale = 0.25] {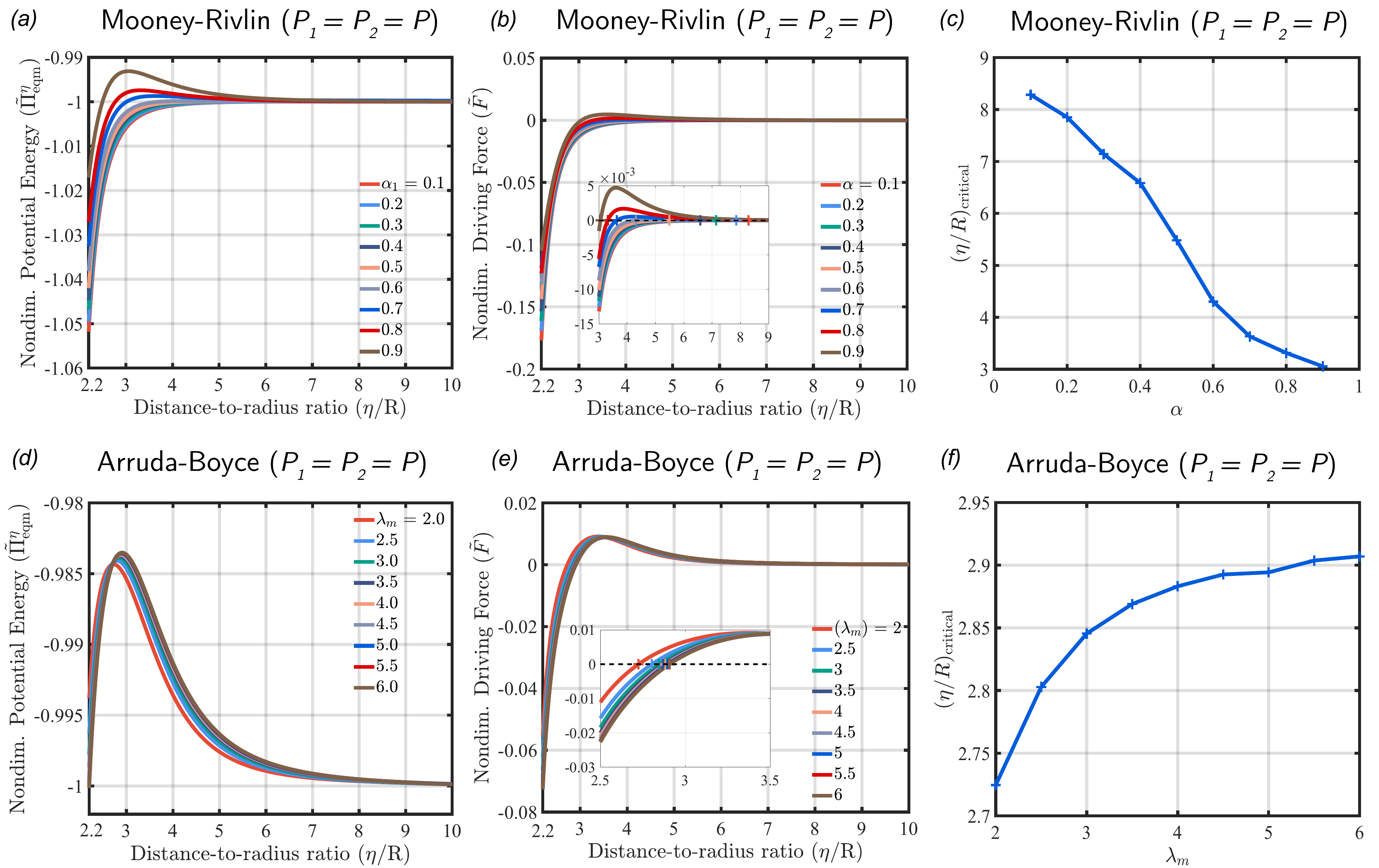}
\caption{\textbf{Effect of strain stiffening.} Comparison of potential energy (a, d), driving force (b, e), and $(\eta/R)_{\rm{critical}}$ (c, f) for different strain-stiffening parameters ($P/\mu=1.5$) for the case of equal positive pressures. For all models a value of $\mu=20$ was chosen for ABAQUS simulations. }
\label{fig:PE_DF_comp_SS} 
\end{figure}

\noindent 
Among our chosen models, Mooney-Rivlin and Arruda-Boyce models can capture strain-stiffening behavior.
The parameters $\alpha$ for Mooney-Rivlin and $\lambda$ for Arruda-Boyce models control the stiffening.
The effect of strain-stiffening on the equal positive pressure case is of particular interest since it can change $(\eta/R)_{\rm{critical}}$.
Figure \ref{fig:PE_DF_comp_SS} shows the variation of potential energy, driving force, and  $(\eta/R)_{\rm{critical}}$ as a function of strain-stiffening parameters for a representative pressure of $P/\mu = 1.5$.

In Mooney-Rivlin model the critical distance is inversely proportional to the strain-stiffening parameter $\alpha$; in Arruda-Boyce model, it is directly proportional to the strain-stiffening parameter $\lambda_m$ but the dependence is not as strong as it is in the Mooney-Rivlin model. 
The parameter $\alpha$ in Mooney-Rivlin model can shift the critical distance significantly (for $\alpha$=1, Mooney-Rivlin model is the same as neo-Hookean model). 
It is worth noting that this behavior is in contrast to that of pressurized cylindrical cavities, where $(\eta/R)_{\rm{critical}}$ is independent of $\alpha$ in Mooney-Rivlin model and strongly dependent on $\lambda$ in Arruda-Boyce model. \\

\subsection{Unequal Pressures}
\begin{figure}
\centering
    \includegraphics[scale = 0.25] {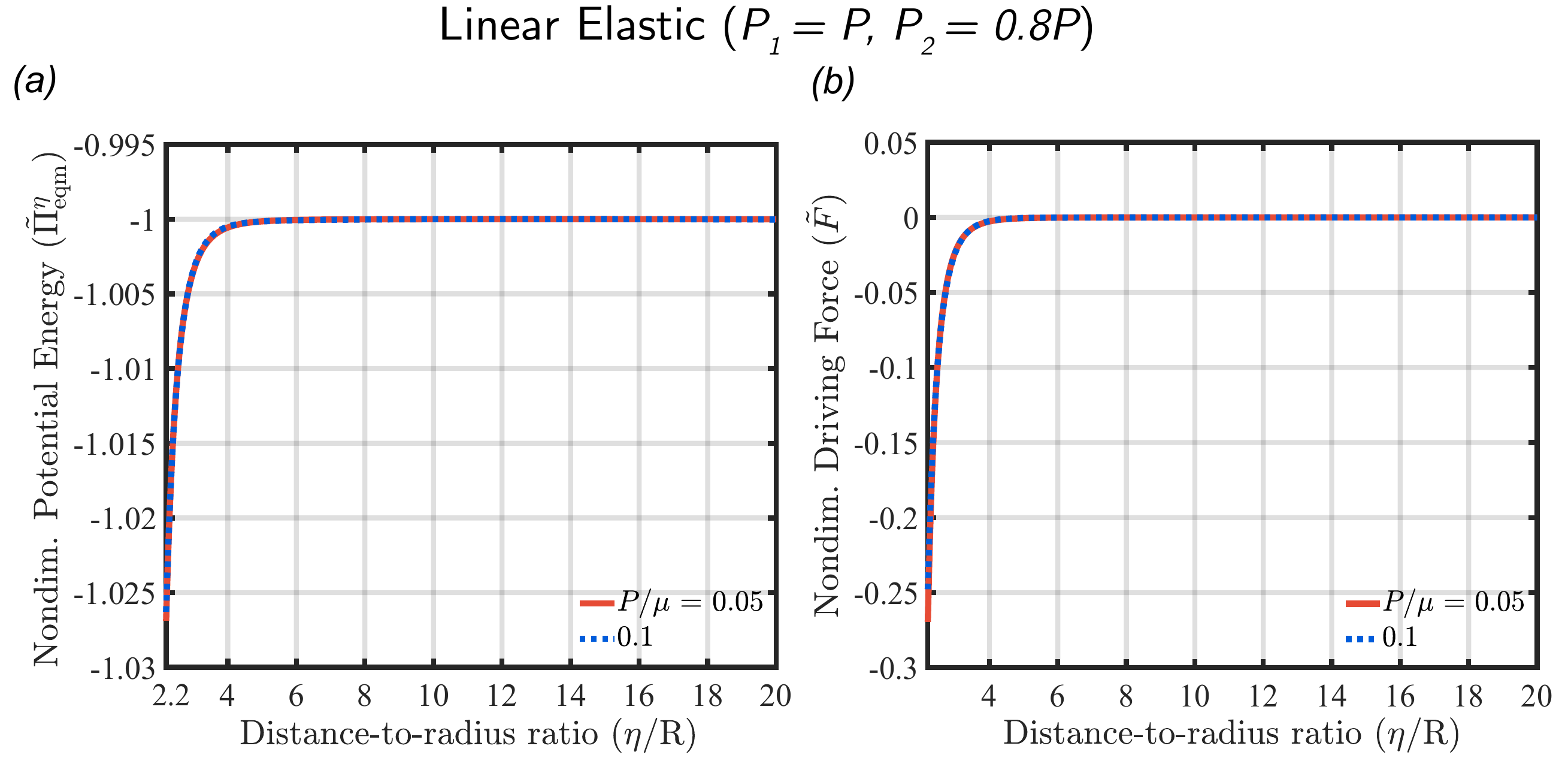}
\caption{Potential energy (a) and driving force (b) for unequal pressures in linear elastic medium. $\mu=20$ was chosen for ABAQUS simulations.}
\label{LE_unequal}
\end{figure}

\noindent To further explore if the intriguing behavior in the case of equal positive pressures is a degenerate case or example of a broader behavior, we carry out simulations for two unequal pressures $P_1\neq P_2$, choosing $P_1 = P$ and $P_2 = 0.8P$.
For a linear elastic medium, we find that the interaction is still always attractive, as seen in Fig.~\ref{LE_unequal}.

\begin{figure}
\centering
    \includegraphics[scale = 0.25] {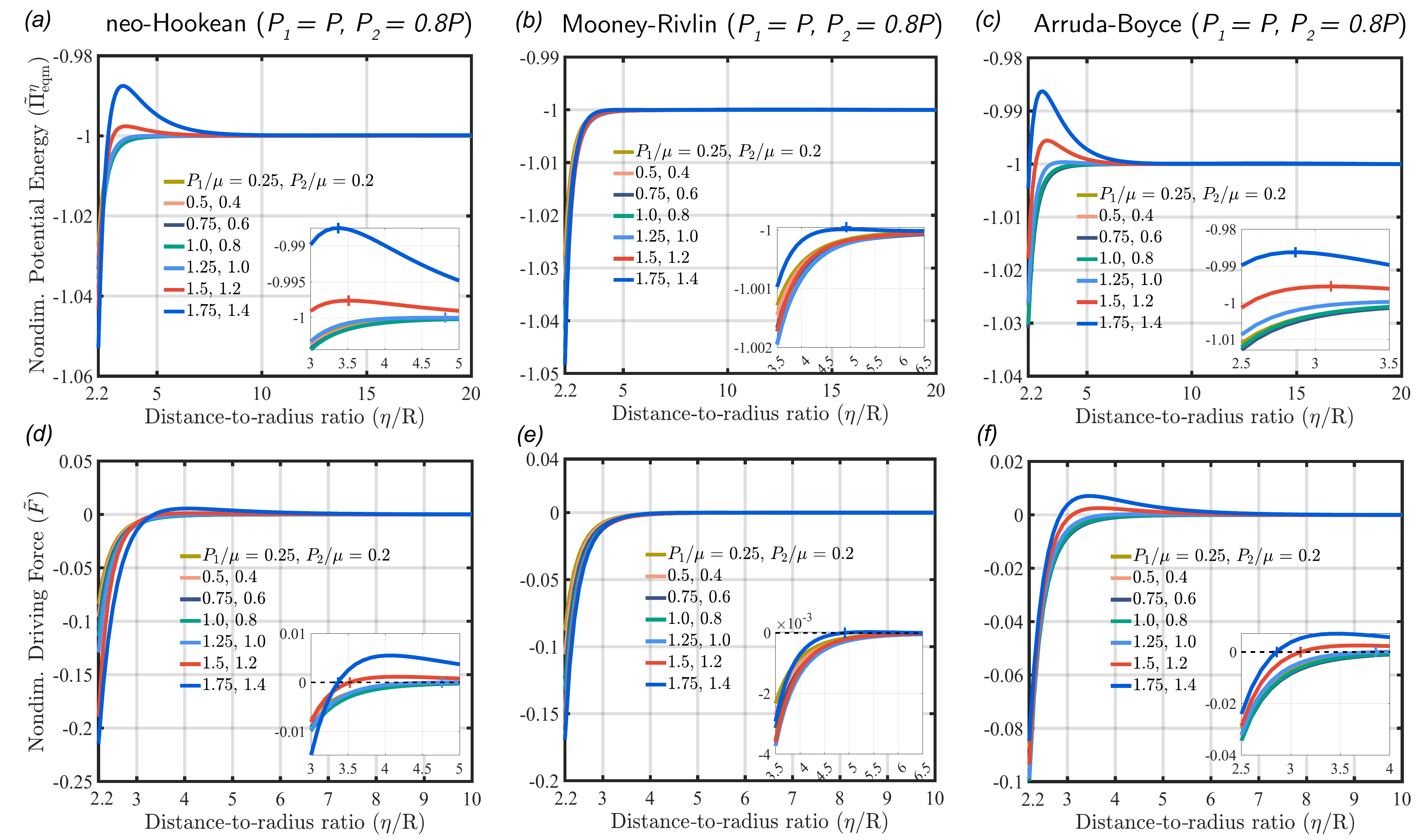}
\caption{\textbf{Unequal positive pressure case.} Potential energy (a-c) and driving force (d-f) for neo-Hookean, Mooney-Rivlin, and Arruda-Boyce models. For all models, a value of $\mu=20$ was chosen for ABAQUS simulations. For Mooney-Rivlin, $\alpha = 0.5$, and for Arruda-Boyce, $\lambda_m=2$, were chosen as the constitutive model parameters.}
\label{fig:PE_unequal}
\end{figure}%
For our chosen hyperelastic material models, the trend in fact remains the same as that of equal positive pressures: there is a critical value of $\eta/R$ at which the interaction turns from attractive to repulsive.
The choice of $P$ and $0.8P$ was made so that both pressures can be in the high pressure range together within the pressure range considered in this study.
The strength of attraction and repulsion is reduced compared to equal positive pressures because of the lower magnitude of loading.

\begin{figure}
\centering
       \includegraphics[scale = 0.25] {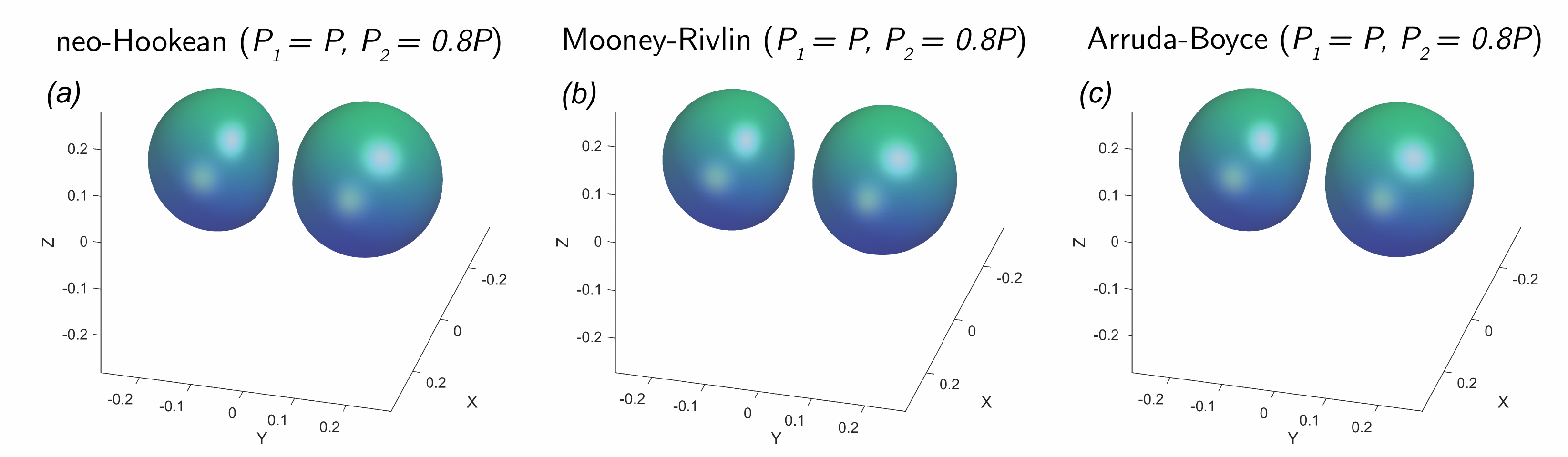}
\caption{Deformed shape of cavities for $\eta/R$=2.8 and $P/\mu$=1.0 with unequal positive pressures for neo-Hookean (a), Mooney-Rivlin (b), and Arruda-Boyce (c) models. For all models, a value of $\mu=20$ was chosen for ABAQUS simulations. For Mooney-Rivlin, $\alpha = 0.5$, and for Arruda-Boyce, $\lambda_m=2$, were chosen as the constitutive model parameters.}
\label{fig:def_crit_III_IV}
\end{figure}

Deformed shape of cavities for unequal pressure loading in separation of $\eta/R$=2.8 and pressure of $P/\mu$=1.0 are shown in Fig.\ref{fig:def_crit_III_IV}.

\section{Conclusion}
\noindent We find that the elastic interaction between two equi-sized pressurized spherical cavities in an infinite elastic matrix is always attractive at low pressures ($P/\mu \lesssim 1.0$) for all linear and nonlinear material models (neo-Hookean, Mooney-Rivlin, and Arruda-Boyce). Non-linear elasticity switches the interaction at higher pressures ($P/\mu \gtrsim 1.0$) only for positive pressures (both equal and unequal), where the potential energy becomes non-monotonic with inter-cavity separation. 
For equal negative pressures the interaction remains attractive even at higher pressures. 
For positive equal and unequal pressures, and for distances smaller than a critical inter-cavity distance, the driving force is negative, representing attractive interaction between the cavities; for separations larger than this critical inter-cavity distance, driving force is positive showing the repulsive nature of the interaction. 
The energy landscape for Mooney-Rivlin and Arruda-Boyce cases show dependence on strain-stiffening parameters $\alpha$ and $\lambda_m$. 
The critical distance in Mooney-Rivlin case is inversely proportional to its strain-stiffening parameter $\alpha$, while in Arruda-Boyce case, increasing the strain-stiffening parameter $\lambda_m$ increases the critical separation.

It is important to note here that the elastic driving forces and the resulting attraction/repulsion only refers to the \textit{tendency} of the cavities to move or `evolve' in a different configuration.
In any physical system, there must be an associated mechanism for the cavities to migrate to be able to observe the effects of driving forces. 
Furthermore, in such systems, there may be numerous other effects at play including surface tension, thermal effects, and chemical effects, which may augment or diminish elastic driving force among cavities.
In light of this understanding, the current work provides new insights into the role of elastic driving forces in complex systems.
Future works could encompass extensions, such as interaction of multiple cavities, effects of material anisotropy and time-dependent elastic properties, and integrate these findings into comprehensive multiphysics modeling frameworks, thereby broadening the applicability of the results to more complex systems.

\section{Acknowledgement}
\noindent 
Support from NH BioMade under the National Science Foundation
EPSCoR award \#1757371 is acknowledged.
Insightful discussions with Prof. Haneesh Kesari (Brown University) and Dr. Chockalingam Senthilnathan (Purdue University) are acknowledged.

\section{Conflict of Interest}
\noindent There are no conflicts of interest.
\newpage
\section*{APPENDIX}
\appendix


\section{Single cavity growth within incompressible neo-Hookean medium: analytical solution}
\label{appendix:single_cavity}
We consider a cavity of radius $A$ in an infinite neo-Hookean medium.
The cavity is internally pressurized with a pressure $P$.
Points in reference coordinate system are denoted by (\textit{R},$\mathrm{\Theta}$,$\mathrm{\Phi}$) and map to $\left(r,\theta,\phi\right)$ in deformed coordinate system, as is usual in a spherical coordinate system.
This problem is spherically symmetric.

The relevant boundary conditions are $\sigma_{rr}(r=a)=-P$ and $\sigma_{rr}(r\rightarrow\infty)=0$ ($\lambda=1$), where $\bfsigma$ is Cauchy stress tensor and $a$ is the deformed radius of the spherical cavity.

The relation between internal pressure, $P$, and the stretch at the internal surface, $\lambda_a$, is \cite{bower2009applied},
{\begin{equation}
    P=\frac{\mu}{2}\Big(1-\lambda_a^{-4}\Big)+2\mu\Big(1-\lambda_a^{-1}\Big),
    \label{eq:P_lambda}
\end{equation}}
where $\lambda_a = a/A$.

We use this equation to determine strain energy, $\Psi_0$, as
{\begin{equation}
    \Psi_0 = 4\pi\mu a^3 \Big(\frac{5}{6}+\frac{1}{2}\lambda_a^{-4}-\frac{1}{3}\lambda_a^{-3}-\lambda_a^{-1}\Big).
\end{equation}}
The potential energy of external forces is
{\begin{equation}
    -P \Delta V=-P\Big[\frac{4}{3}\pi(a^3-A^3)\Big]=-\frac{4}{3}P\pi A^3 (\lambda_a^3 - 1).
\end{equation}}
Finally, potential energy of the system, $\Pi_0$, is calculated as
{\begin{equation}
    \Pi_0 = \Psi_0 - P \Delta V = 4\pi\mu a^3 \Big(\frac{5}{6}+\frac{1}{2}\lambda_a^{-4}-\frac{1}{3}\lambda_a^{-3}-\lambda_a^{-1}\Big) -\frac{4}{3}P\pi A^3 (\lambda_a^3 - 1)
    \label{eq:PE_lambda}
\end{equation}} 

\begin{figure}[H]
\centering
    \includegraphics[scale = 0.25] {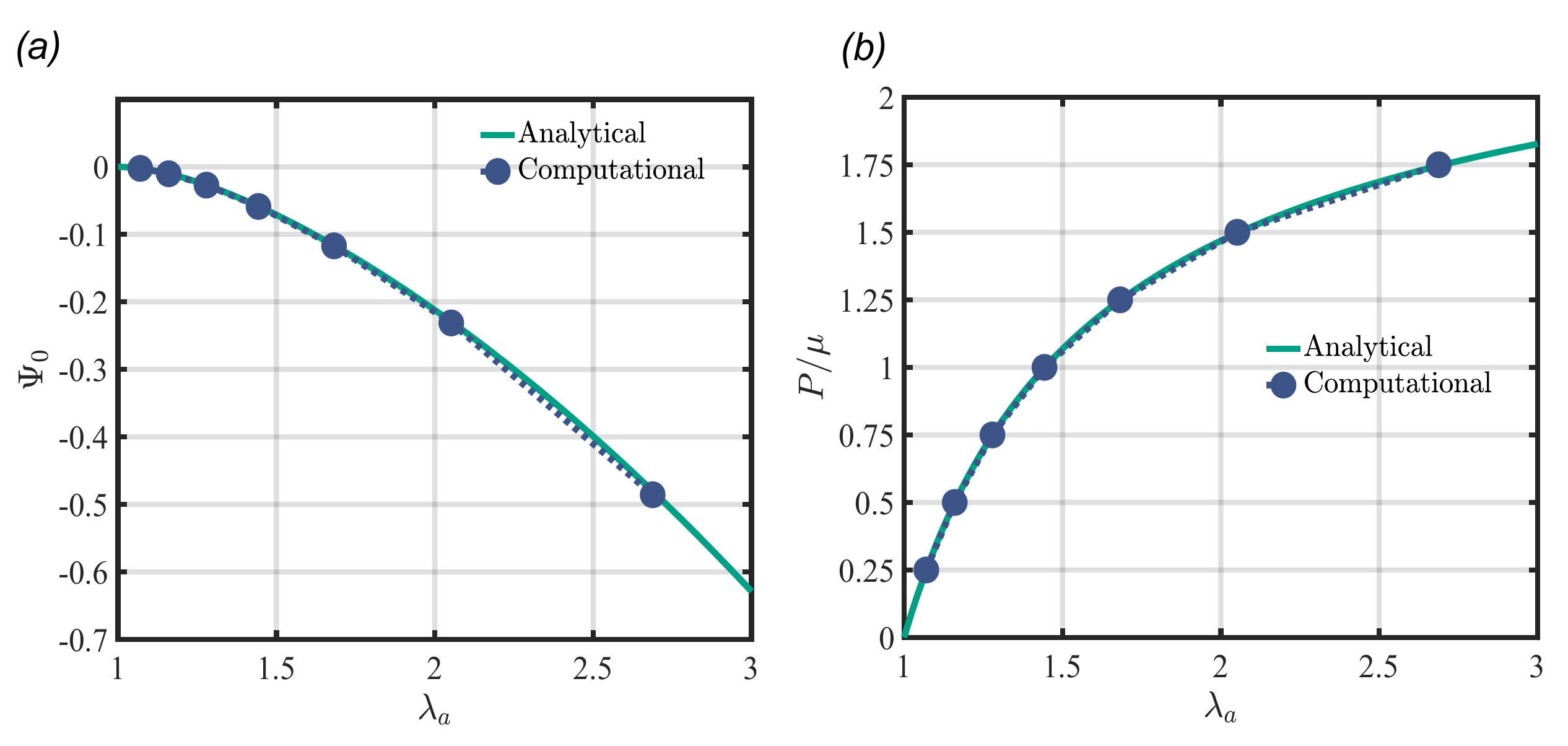}
\caption{Potential energy (a) and pressure (b) as a function of radial stretch for a single cavity within a neo-Hookean medium.}
\label{fig:single_hole}
\end{figure}

The computational solution of a single cavity embedded in a neo-Hookean medium is verified using Eq. \ref{eq:P_lambda} and Eq. \ref{eq:PE_lambda} and shown in Fig. \ref{fig:single_hole}.

\section{Actual data before fitting}
Original potential energy plots are shown below; they have been fitted with a smooth curve to yield a smooth first derivative and avoid introducing noise into the interaction energy.

\begin{figure}[H]
\centering
    \includegraphics[scale = 0.25] {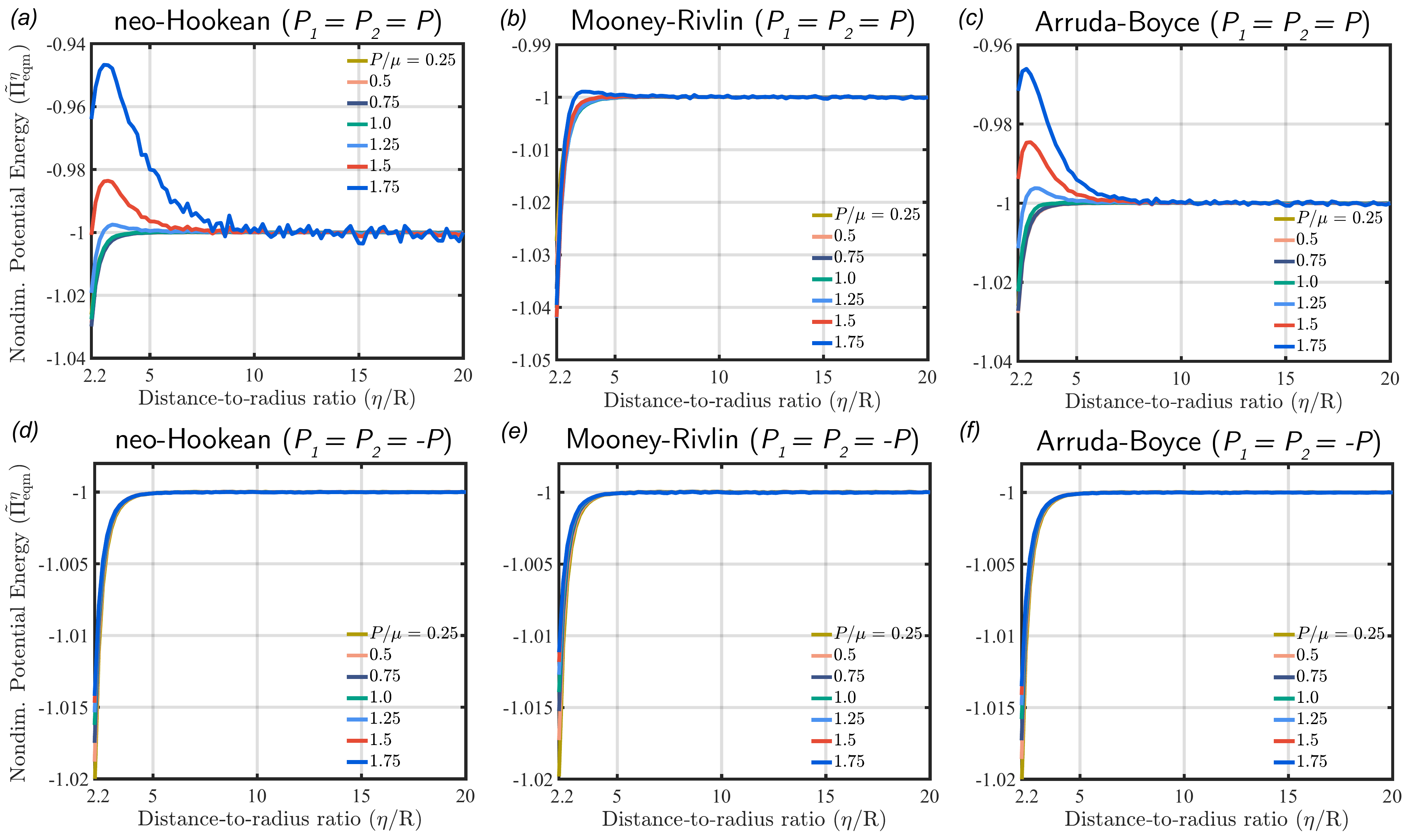}
\caption{Original potential energy of equal positive (a-c) and equal negative (d-f) pressures for neo-Hookean (a,d), Mooney-Rivlin (b,e), and Arruda-Boyce (c,f) models. For all models, a value of $\mu=20$ was chosen for ABAQUS simulations. For Mooney-Rivlin, $\alpha = 0.5$, and for Arruda-Boyce, $\lambda_m=2$, were chosen as the constitutive model parameters.}
\label{fig:PE_equal_initial}
\end{figure}%

\begin{figure}[H]
\centering
    \includegraphics[scale = 0.25] {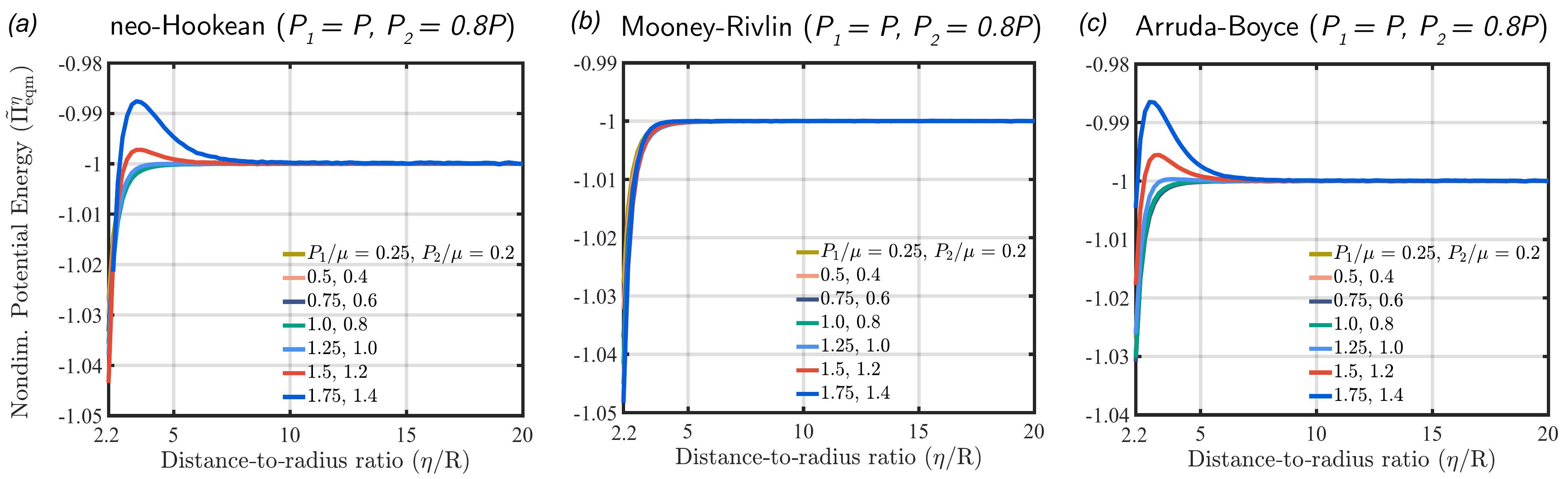}
\caption{Original potential energy of unequal pressures for neo-Hookean (a), Mooney-Rivlin (b), and Arruda-Boyce (c) models. For all models, a value of $\mu=20$ was chosen for ABAQUS simulations. For Mooney-Rivlin, $\alpha = 0.5$, and for Arruda-Boyce, $\lambda_m=2$, were chosen as the constitutive model parameters.}
\label{fig:PE_unequal}
\end{figure}%

\section{Mesh-independence study}
\begin{figure}[H]
\centering
    \includegraphics[scale = 0.25] {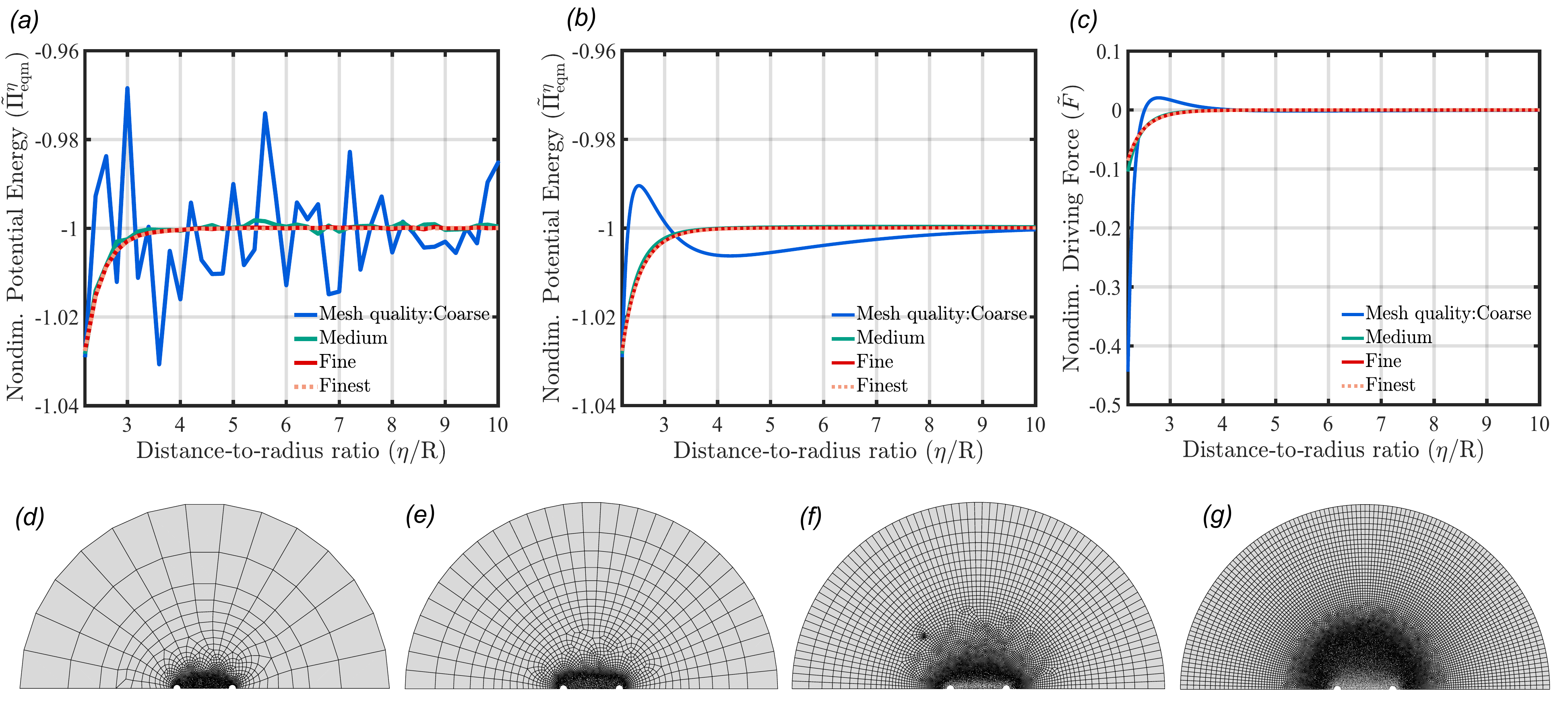}
\caption{Mesh-independence study for a constant distance ($\eta/R=15$) and a neo-Hookean model ($P/\mu=1$) (a) initial potential energy, (b) fitted potential energy, (c) driving force, with mesh qualities shown as (d) coarse, (e) medium, (f) fine, (g) finest.}
\label{fig:mesh_converg}
\end{figure}

We have carried out a mesh convergence study to ensure that the results are mesh-independent. The potential energy and driving force plots are depicted in Fig.\ref{fig:mesh_converg} for an equal pressure case with $P/\mu=1$ and $\eta/R=15$. Among four different mesh qualities, the `Fine' mesh provides good accuracy and we have used it in this paper.

 \bibliographystyle{elsarticle-num}
 \bibliography{cas-refs}

\end{document}